# Carbon chains and graphene nucleus synthesized on Ni(111) surface

T. V. Pavlova[1,2,][*], S. L. Kovalenko[1], K. N. Eltsov[1,2]

[1] Prokhorov General Physics Institute, Russian Academy of Sciences, ulitsa Vavilova 38, 119991 Moscow, Russia

[2] Faculty of Physics, National Research University Higher School of Economics, ulitsa Myasnitskaya 20, Moscow 101000, Russia

**Abstract**

Linear chains of about 10-13 carbon atoms were predicted to be the most favorable phase on different metal surfaces prior to graphene nucleation. However, unlike the graphene that widely studied both theoretically and experimentally, carbon chains on metal surfaces were not directly studied by STM yet. Here we fill in the gap and report on STM experiments of linear carbon chains synthesized on Ni(111) through on-surface coupling of dehydrogenated propene molecules. Identification of chains was supported with DFT calculations and the proposed models consist of 12 carbon atoms, possibly covered by hydrogen atoms. Heating to 580 K leads to dramatic decrease of carbon chains and new phase appearance – graphene nucleus coexisted with nickel carbide. After flash annealing to 773 K (temperature of graphene synthesis), a small number of chains were presented on the Ni(111) surface, together with graphene islands and nickel carbide. The carbon chains are stable at room temperature and their mobility was directly observed by STM.

[*] Corresponding author: Tel: +7 499 503 8784.
E-mail address: pavlova@kapella.gpi.ru (T. Pavlova)

# 1. Introduction

One-dimensional monoatomic chains of carbon atoms has been less investigated than two- and three-dimensional carbon allotropic forms, but in the last decades carbon chains attracted more attention because of many interesting properties [1,2] and potential applications. Carbon chains have been proposed for developing energy storage devices and as a part of organic electronics due to its tunable electrical properties [1]. Not only infinite or very long linear chains are of interest but also ten-atom carbyne chain can be utilized, for example, as a connector between two graphene sheets in nanoelectronics [3]. The synthesis of carbon chains has remained a big experimental challenge because they are more reactive and unstable than 2D and 3D carbon structures. In contrast, graphene nucleus are much easy to create and it was done on Cu, Ru, Rh, and Ir metal supports, theoretical and experimental studies were summarized in review [4]. Small graphene islands are also of great importance for nanoelectronics [5,6].

Monoatomic carbon chains can be stabilized on the metals surfaces and thereby it is possible to study them with scanning tunneling microscope (STM). To the best of our knowledge, some disordered carbon clusters were observed on Cu [7] and Ru(0001) [8], however, carbon chains on metals have not been studied with STM. To investigate properties of 1D carbon chains on metals, first principles modeling have been used: the formation energy [9], electronic and geometric structures [10], mobility, thermodynamic stability, and possibility of decomposition [11] are systemically explored. Although the interactions between carbon and various metals are different, the lowest-energy configuration of less than 10-13 carbon atoms was predicted to be a linear chain on different metals (Cu, Ni, Rh, and Ru) [9, 10]. At the number of C atoms increasing, the two-dimensional carbon network becomes more stable and the phase transition from carbon chain to graphene should occurs [9, 10].

Inspired by recent theoretical investigations of carbon chains on metal surfaces and prediction that the chains are the phase of carbon prior to graphene, we have synthesized them on nickel. Ni(111) substrate was chosen because it has strong catalytic properties for hydrocarbon molecules which has been studied both theoretically [12-14] and experimentally [15, 16]. Nickel has also been used extensively to grow graphene because graphene forms nearly a co-lattice with Ni(111) surface. For the first time, we obtained carbon chains on Ni(111) during the temperature programmed growth of graphene [17], and we have studied them in depth in the present work.

The objective of our work is the carbon chains synthesis by propene adsorption on Ni(111) and investigation of them with a STM in combination with density functional theory (DFT) calculations for an identification of the observed structures. To study the phase transition from carbon chains to graphene nucleus, the substrate was annealed up to temperature of carbide (580 K) and graphene (773 K) formation. The stability and mobility of carbon chains were directly observed. Two models of carbon chains with and without hydrogen were proposed to explain synthesized structures.

## 2. Methods

### 2.1. Experimental method

All experiments were performed in an ultra-high vacuum (UHV) system. The system is equipped with a Riber OPC-200 electron Auger spectrometer, a Riber OPC-304 low energy electron diffractometer (LEED), and a GPI 300 scanning tunneling microscope [18]. The clean Ni(111) surface of the (6×6×2 mm) single crystal was prepared by cycles of ion etching (ion energy 1 keV) for 15 min and subsequent annealing at 873 K for 10 min. Cycles of sample cleaning from the dissolved carbon include annealing at 573 K (optimal temperature for carbon segregation to the surface [19]) for 15 min followed by ion etching (ion energy 1 keV) for 15 min. The quality of the surface was monitored by the measurements with STM, LEED and electron Auger spectrometer. The adsorption of propene was carried out from the effusion beam at a distance of about 2 cm from the Ni(111) surface. The gas pressure in the beam was controlled by a piezoceramic leak valve and could vary from $10^{-9}$ to $10^{-5}$ Torr. STM-images were recorded at room temperature, polycrystalline tungsten tips were used.

### 2.2. Computational methods

Spin-polarized DFT calculations were performed using the generalized-gradient approximation (GGA) according to Perdew, Burke, and Ernzerhof (PBE) [20] and the projector-augmented-wave (PAW) methods [21] implemented into VASP code [22]. Semiempirical Grimme's DFT-D2 dispersion correction [23] was applied for all calculation. Two slabs of 6x6 and 8x6 were built to represent the substrates for short ($C_3H_x$) and long chains ($C_{12}H_x$), respectively. The slab 6x6 (8x6) was four (three) layers thick. The two lowest layers were fixed to their bulk positions while all the other atoms were allowed to relax. A vacuum space of 15 Å is introduced to isolate the top of one slab from the bottom of the next slab. STM images were simulated using Hive program [24] in the framework of Tersoff−Hamann approximation [25] considering states between Fermi energy ($E_F$) and $E_F - 0.5$ eV.

Adsorption energy was calculated as

$$E_{ads} = E_{tot} - E_{slab} - E_{mol}, \quad (1)$$

where $E_{tot}$ corresponds the energy of an adsorbate–surface system, $E_{slab}$ stands for the energy of a clean Ni(111) slab, $E_{mol}$ is the energy of a propene in vacuum.

## 3. Results and discussion

### 3.1. Experimental results

Carbon chains were synthesized by the exposure of the clean Ni(111) to propene ($C_3H_6$) at room temperature. Figure 1 shows two STM-images of the Ni(111) surface after two different propene doses, 0.5 L and 300 L. At adsorption doses greater than 300 L the concentration of carbon structures on the surface did not increase, so in Fig. 1b the saturated structure is given. We have confirmed by using electron Auger spectrometer that there are carbon structures on surface. We have interpreted the bright elongated objects on STM-images as carbon chains. Some of the chains are linear, others are curved, with an average length of about 20 Å.

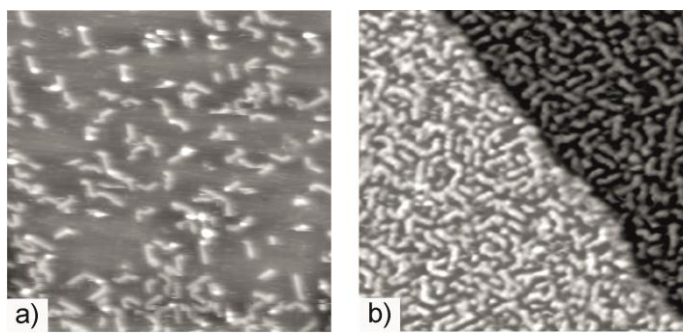

Fig. 1. STM-images (275×275 Å², U = −1311 mV, I = 0.2 nA) of Ni(111) after exposure to 0.5 L (a) and 300 L (b) propene at room temperature.

Ni(111) sample with adsorbed propylene (25 L) was heated to a temperature just above and below the temperature of the C-C bonds breaking in the hydrocarbons ($C_2H_4$ and $C_2H_2$) on Ni(111) [16]. After annealing to 573 K for 10 min, the number of carbon chains on the surface decreased significantly (Fig. 2a). New carbon clusters were observed which can be attributed to graphene nucleus. The minimum size of the nucleus is about 12 Å. The graphene structure was resolved on a small island, adjacent to the carbon chain (see insert in Fig. 2a). With further heating to 623 K, the graphene islands increased (Fig. 2b). The described structures coexist with nickel carbide, which most effectively grows at these temperatures.

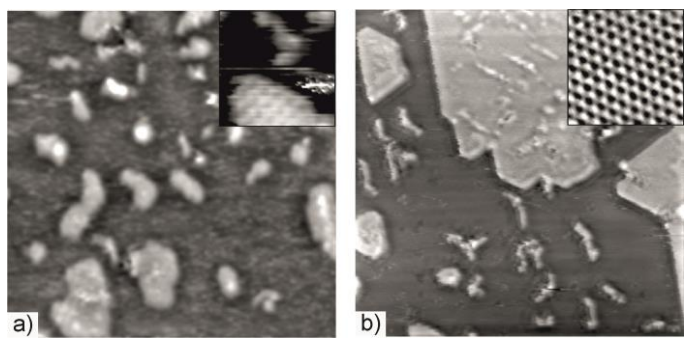

Fig. 2. a) STM-image (274×274 Å$^2$, U = −1311 mV, I = 0.1 nA) of Ni(111) surface exposed to 25 L propene at room temperature and annealed at 573 K for 10 min, b) STM-image (274×274 Å$^2$, U = −64 mV, I = 0.5 nA) of Ni(111) surface exposed to 25 L propene at room temperature and annealed at 573 K for 40 min then 623 K for 60 min. STM-images (22×22 Å$^2$) on inserts show graphene structure of small islands.

To study the transformation of Ni(111) covered by carbon chains at the temperature of graphene synthesis, the sample was annealed to 773 K for 5 minutes. After cooling the sample to room temperature, following structures were observed on STM-images: graphene islands on terraces and steps, nickel carbide near the steps, and carbon chains on terraces (Fig. 3). The concentration of chains decrease significantly compared to the concentration before the sample heating (Fig. 1). The smallest graphene island observed after to 773 K has a triangle side of the order of 50 Å.

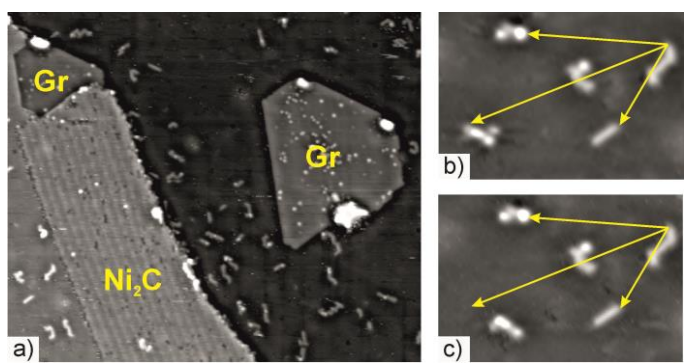

Fig. 3. a) STM-image (688×592 Å$^2$, U = −815 mV, I = 0.2 nA) of Ni(111) surface exposed to 500 L propene at room temperature and annealed at 773 K for 5 min. Carbide and small graphene island are formed at step, bigger graphene island is formed on terrace. b-c) STM-images (160×110 Å$^2$, U = −815 mV, I = 0.2 nA) of carbon chains on Ni(111) after annealing at 773 K for 5 min. STM-image c) was recorded after 4 min after b).

The mobility of chains is envisaged in Fig. 3b, where the motion of the single carbon chain is shown. Two STM-images were recovered at room temperature with delay of 4 min. During this time, the chain was shifted by 20 Å. Thus, we have observed that the chains are stable at room temperature.

After several cycles of sample cleaning, described in Section 2.1, carbon chains were observed on the Ni(111) surface (Fig. 4a) without propene adsorption. However, if the sample is cleaned for a long time (many cycles), it is possible to obtain a clean surface with very small chains concentration (see Fig. S1 in Supplementary material). It is difficult to obtain STM-image of carbon chains with high resolution, but the average size can be determined as 20 Å (Fig. 4b).

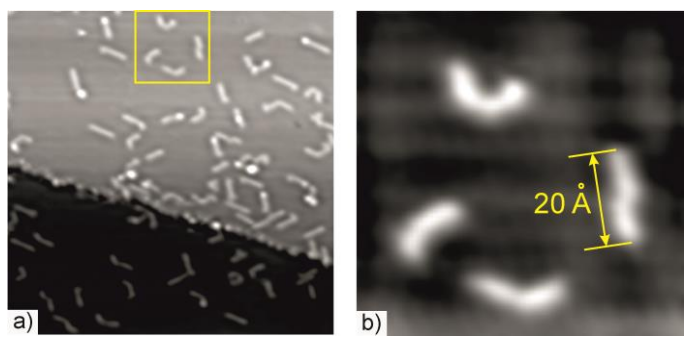

Fig. 4. a) STM-image (275×275 Å$^2$, U = -1311 mV, I = 0.2 nA) of Ni(111) surface after cleaning procedure described in section 2.1, b) STM-image (67×67 Å$^2$) of Ni(111) surface with carbon chains shown by square in a).

## 3.2. Theoretical results

An adsorbed propene molecule is tilted at an angle 26º to Ni(111) surface. The lowest carbon atom of the adsorbed $C_3H_6$ molecule (bounded to two H atoms) is located in the three-fold position at distances of 2.11-2.20 Å to the three nearest nickel atoms (Fig. 5a). The middle C atom (bounded to one H) is situated on top of the Ni atom with bond length d (C-Ni) = 1.99 Å. The upper C atom (bounded to three H) is in the three-fold position at the largest distance from the three Ni atoms 2.98 - 3.22 Å. The calculated length of the double bond (C=C) in free propene molecule is 1.34 Å, while this bond is elongated to 1.45 Å on nickel substrate due to interaction of C atoms with Ni. The single bond (C-C) of the propene molecule does not change after adsorption (1.51 Å).

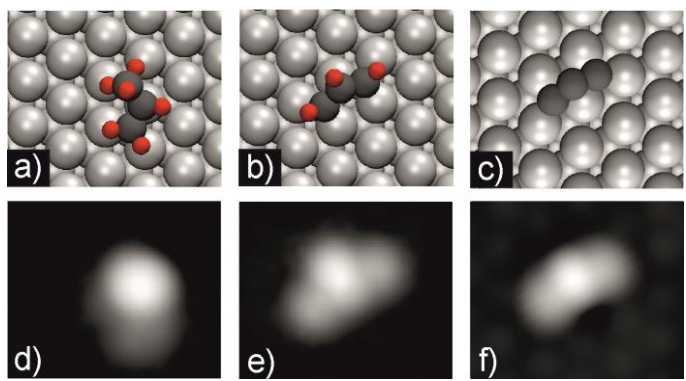

Fig. 5. Structural models and theoretical STM-images of $C_3H_6$ (a, d), $C_3H_3$ (b, e), and $C_3$ (c, f). Nickel are represented by large light-gray atoms, carbon by small gray atoms and hydrogen by the smallest red atoms.

Nickel is known for its catalytic properties with respect to hydrocarbons, so the propene molecule adsorbed on Ni(111) can partially or completely dehydrogenate. In particular, ethylene ($C_2H_4$) adsorbed on Ni(111) dehydrogenates to acetylene ($C_2H_2$) in the temperature range 200 - 230 K [15]. This suggests that the dehydrogenation of the propene molecule $C_3H_6 \rightarrow C_3H_x$ + (6-x)$H_{ad}$ can occur on Ni(111) surface below room temperature.

We have modeled several configurations with three carbon and six hydrogen atoms to find the most energetically preferable surface structure. We have started from propene molecule on Ni(111), followed by deleting of three and six hydrogen atoms from $C_3H_6$ through the addition of H atoms onto the surface. The last considered configuration was a fully dissociated molecule up to three adsorbed carbon ($C_{ad}$) and six hydrogen ($H_{ad}$) atoms. Adsorbed hydrogen atoms ($H_{ad}$) were placed at the face-centered-cubic (fcc) hollow sites as the most preferable ones on Ni(111) surface [26]. In partially dehydrogenated structure, $C_3H_3$, both C-C bonds are equal to 1.47 Å, whereas in completely dehydrogenated carbon chain, $C_3$, C-C bonds are shortened to 1.35 Å. Distanced to the nearest Ni atoms are vary from 1.96 to 2.12 Å for $C_3H_3$ (Fig. 5b) and from 1.82 to 2.03 Å for $C_3$ (Fig. 5c). Further, both $C_3$ and $C_3H_3$ will be called a carbon chain covered by hydrogen or not.

For all species studied, the adsorption energies decrease in the order: $E_{ads}(C_3H_3+3H_{ad})$ = −1.90 eV < $E_{ads}(C_3H_6)$ = −1.69 eV < $E_{ads}(C_3+6H_{ad})$ = −1.38 eV <  $E_{ads}(3C_{ad}+6H_{ad})$ = −0.41 eV. The most preferred structure of the four examined is $C_3H_3$ + 3$H_{ad}$. Although we did not test how many hydrogen atoms are attached to the $C_3$ chain, we can conclude that the partially dehydrogenated propene molecule ($C_3H_3$) is the most energetically preferable state on Ni(111). This is consistent with the finding that the most favorable state of $C_2H_4$ on Ni (111) is $C_2H_2$ + 2$H_{ad}$, i.e. partially dehydrogenated ethylene molecule (at coverage < 0.4 ML) [14, 15].

Calculated STM-images for the optimized structures for $C_3H_6$, $C_3H_3$, and $C_3$ are shown on Fig. 5. The longitudinal dimension of $C_3H_3$ and $C_3$ chains is approximately the same and is about 6 Å. The experimental STM-images, though, obviously show a longer linear chains with length of about 20 Å. This observation lead to the assumption that the chains are formed from several fragments of the adsorbed propene. Indeed, chains of three carbon atoms are very mobile on the Ni(111) surface and the minimal diffusion barrier (0.21 eV) is two times lower than the barrier of a single carbon atom diffusion over the surface (0.48 eV) [11]. The reaction of two $C_3$ chains coupling to one $C_6$ is energetically favorable with the activation barrier of 1.22 eV [11]. The activation barrier is comparable with the barrier of dehydrogenation (about 1 eV for the reaction $C_2H_4 \rightarrow C_2H_2$ + 2$H_{ad}$ on Ni (111) [13]) and, thus, can be overcome at room temperature. Moreover, theoretical calculations of various cluster structures up to 24 carbon atoms on Ni(111) [9] found the linear chains being the most stable structure at the number of C atoms less than 12.

We have constructed a chain of 12 carbon atoms as a model of four joined $C_3$ fragment, and, keeping in mind the lowest-energy structure ($C_3H_3$), we have also built a chain $C_{12}H_{12}$ (Fig. 6a,b). The C-C bonds in the $C_{12}$ chain adsorbed on Ni(111) are lengthened (1.33-1.36 Å) relative to a carbon chain in the gas phase (1.29 Å). Owing to the closer distance between the surface Ni atoms and edge C atoms (about 1.76 Å) compared to the other carbon atoms (about 2.0 Å), edge atoms bind more strongly with surface. The angles between the C atoms in the chain vary from 143° at the ends of the chain to 160° in the middle, so the carbon chain on metal surface is not perfectly linear as in vacuum. In the $C_{12}H_{12}$ chain, the bond lengths between the carbon atoms are approximately the same along the entire chain and equal to 1.44 Å; the distances from the C atom to the nearest Ni atom ranges from 1.89 Å at the edges to 2.0 Å in the center of chain (Fig. 6b). The angles between the carbon atoms are about 120-126°, thus the hydrogen covered carbon chain is more curved than the one without hydrogen.

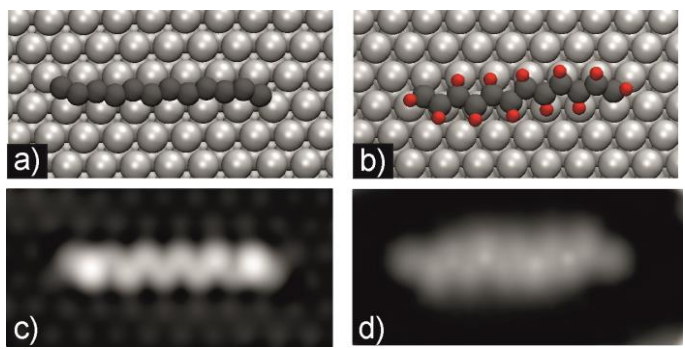

Fig. 3. Structural models and theoretical STM-images of long chains $C_{12}$ (a, c) and $C_{12}H_{12}$ (b, d). Nickel are represented by large light-gray atoms, carbon by small gray atoms and hydrogen by the smallest red atoms.

To verify that the long carbon chains are observed on Ni(111), we have calculated STM-images (Fig. 6c,d). The electronic density distribution along $C_{12}$ and $C_{12}H_{12}$ chains is about 17 Å, that a bit shorter than experimentally observed (about 20 Å). Note that the transverse size of the chains is also smaller on the theoretical STM-image (~ 3.0 Å) compared with the experimental one (~ 5.5 Å). If we assume that the STM tip blurs the electron density in the longitudinal and transverse directions equally, the length of chains on calculated STM-images will be increased from 17 to 19.5 Å. Thus, we can conclude that the propene adsorption leads to the formation of carbon chains consisting on average of four $C_3$ fragments, 12 carbon atoms with or without hydrogen. Because the chains are not imaged with high resolution, it is difficult to prove experimentally which of the two chains (hydrogenated or not) was observed and therefore this question remains unanswered.

It is known that C-C bonds in the carbon chains on Ni (111) are broken at the temperature 773 K, and we refer to the work [16] in which it was shown that ethylene and acetylene on Ni(111) were decomposed completely into individual C and H atoms at a temperature less than 600 K. Carbon atoms from chains can dissolve in the bulk or be involved in the graphene formation. As it was predicted in [3, 4], carbon cluster of more than 12 atoms will form graphene nucleus. Thus, at the temperature increasing up to 573 K carbon atoms segregate from bulk and form graphene nucleus coexisted with carbon chains. Probably, there are two channels for carbon chains formation on Ni(111) after sample annealing: the adsorption of hydrocarbons from gas phase and the segregation of carbon atoms. Segregated atoms which form a chain can be covered by hydrogen adsorbed from the gas phase since there is a small amount of molecular hydrogen in the chamber even under $2 \times 10^{-10}$ Torr.

The important role of atomic steps on the surface in the decomposition of hydrocarbons on Ni(111) was noted in previous works, in particular, it was experimentally established that the presence of steps accelerates the acetylene decomposition, so the molecules dissociate to separate atoms at a temperature of 180 K [15]. This agrees with our observation that there are no chains located along the steps edges (Fig. 4a). Single bright spots near steps edges (Fig. 4a) can be interpreted as carbon atoms at step edge. This cites (between the four Ni atoms: two Ni atoms of step and two of the terrace) are very favorable for carbon atoms [27]. Therefore chains trapped at the step edges can dissociate onto single atoms.

## 4. Summary

Carbon chains were synthesized by the propene adsorption onto the Ni(111) surface. Based on the results of DFT calculations, the atomistic models of the carbon chains were given. Two chains structures were proposed: chain of carbon atoms only and carbon bounded with hydrogen atoms. The mechanism of chains formation is explained by a coupling of several partially ($C_3H_3$) or completely ($C_3$) dehydrogenated fragments of adsorbed propene molecules. The process of $C_{12}$ ($C_{12}H_{12}$) chains formation from $C_3$ ($C_3H_3$) fragments runs at room temperature. After heating to the graphene synthesis temperature (773 K), the concentration of carbon chains on the surface decreased significantly. It was directly observed that the chains are mobile on terraces at room temperature. The dissociation of chains into carbon atoms at the steps edges is clearly seen on STM-images. To determine which of the chains are realized (covered by hydrogen or not, or may be both of them), further structural and spectroscopic analyzes are needed.

**Acknowledgements**

This work was supported by Grant No. 16-12-00050 from the Russian Science Foundation. We are grateful to the Joint Supercomputer Center of RAS for the possibility of using their computational resources for our calculations.

**Supplementary material**

1. The Ni(111) surface can be cleaned so well that the concentration of carbon chains is very small (Fig. S1). To obtain such a surface, many cleansing cycles described in 2.1 should be used.

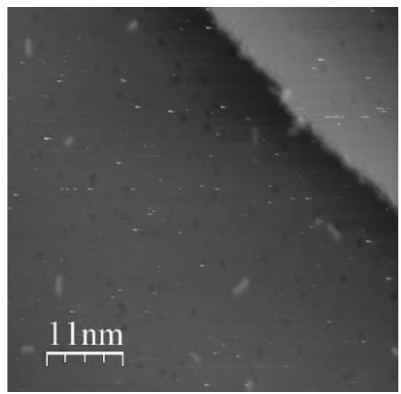

Fig. S1. STM-image (525×525 Å$^2$, U = −464 mV, I = 0.3 nA) of clean Ni(111) surface.

2. Carbon chains can appear on the nickel surface during the scanning process at the high voltage (Fig. S2a). New chains generated by STM tip can be seem on the large-scale STM-image (Fig. S2b).

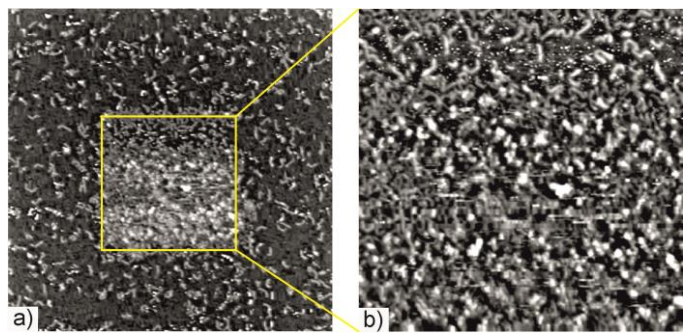

Fig. S25. a) STM-image (491×491 Å$^2$, U = −3525 mV, I = 0.2 nA) of Ni(111) surface after propene adsorption, b) large-scale STM-image (2021×2021 Å$^2$, U = −1734 mV, I = 0.2 nA) of Ni(111) surface after scanning of small area shown in a).